\documentclass[10pt]{article}
\usepackage[OE]{express}

\usepackage{bm}% bold math
\usepackage{xcolor}
\usepackage{commath}
\usepackage{mathrsfs,amsmath}
\usepackage{gensymb}
\usepackage{wrapfig}

\begin{document}
%Common-path phase-shifting digital holography for modal decomposition of vortex beams
\title{Efficient Modal Decomposition of Vortex Beams
via Holographically Reconstructed Phase}

% \author{Jasmine M. Knudsen, Samuel N. Alperin, Andrew A. Voitiv,\\William G. Holtzmann, Juliet Gopinath and Mark E. Siemens\authormark{*}}

% \address{Department of Physics \& Astronomy, University of Denver, Denver CO 80210 USA}

\author{Jasmine M. Knudsen,\authormark{1} Samuel N. Alperin,\authormark{1} Andrew A. Voitiv,\authormark{1} William G. Holtzmann,\authormark{1} Juliet T. Gopinath,\authormark{2,3} and Mark E. Siemens\authormark{1,*} }

\address{\authormark{1}Department of Physics \& Astronomy, University of Denver, Denver, Colorado 80210, USA\\
\authormark{2}Department of Electrical, Computer, and Energy Engineering, University of Colorado, Boulder, Colorado 80309, USA\\
\authormark{3}Department of Physics, University of Colorado, Boulder, Colorado 80309, USA}
\email{\authormark{*msiemens@du.edu}} %% email address is required

\begin{abstract}
We use phase-shifting digital holography to measure the amplitude and phase of twisted light. In our experiment, a spatial light modulator generates the studied vortex beams in addition to a co-propagating reference beam with a controllable relative phase. We show complex field measurements for single and superposition Laguerre Gaussian (LG) modes, demonstrate full modal decompositions into LG and orbital angular momentum (OAM) power spectral bases, provide error analysis and demonstrate high insensitivity to detector misalignment to show the robustness of the technique. This enables rapid determination of OAM spectra with low uncertainty, allowing us to  report the first vortex beams with $99.9\%$ purity.
\end{abstract}

 % REPLACE WITH CORRECT OCIS CODES FOR YOUR ARTICLE, MINIMUM OF TWO; Avoid using the OCIS codes for “General” or “General science” whenever possible.
%For a complete list of OCIS codes, visit: https://www.osapublishing.org/oe/submit/ocis/
\bibliographystyle{unsrt}
\bibliography{PSholoOAM2}

\section{Introduction}
An optical mode with orbital angular momentum (OAM) is characterized by a helical wavefront with an integer winding number $\ell$, leading to an OAM per photon of $\ell \hbar$ \cite{Allen1992}. Light's OAM provides a discrete parameter space, with bounds limited only by the numerical aperture of the system \cite{Chen2016,Restuccia2016}, that has already been utilized in high-torque light-matter interactions \cite{Simpson1997}, quantum entanglement \cite{Fickler2014,Mair2001}, and terabit-bandwidth communications based on multiplexing OAM states. \cite{Bozinovic2013,Huang2014}. While the helical phase of light is an efficient carrier of angular momentum information, it is challenging to measure and characterize because it does not directly affect the intensity of the light measured by a camera.

There are many demonstrated methods to characterize and measure light's OAM. The simplest and oldest method is to interfere the beam with a Gaussian reference; the number of azimuthal interference fringes is equal to $| \ell |$ \cite{Vickers2008}. Diffraction from apertures of various geometries has also been shown to contain information about the nearest-integer $\ell$ \cite{Ambuj2014,Anderson2012}, even in the presence of other modes \cite{jesus2012study}. More recent work has shown that the average OAM can be measured quantitatively with a cylindrical lens\cite{Alperin2016} or by integration of the local OAM density \cite{Leach2004}. Finally, modal decomposition methods can take measurements of the OAM power spectrum by log-polar optical transformations \cite{Berkhout2010} or diffraction forked grating correlation filters \cite{Schulze2013,Litvin2012}. Despite all of these methods, there still exists a need for a complete and efficient way to measure the OAM of an arbitrary beam.

The fundamental challenge of measuring optical OAM lies in measuring the helical phase of OAM carrying modes: if the full, complex field of a mode can be measured, then the beam will be completely characterized. Unfortunately, the two most common methods of measuring a complex mode are insufficient: wavefront sensors are expensive, low resolution, and cannot handle the phase singularities necessarily present within twisted light. Interferometry can recover only partial phase information in the form of a forked grating \cite{Fang2017}, because of sign ambiguity caused by the inverse cosine operation in the recovery of phase information from an interferogram \cite{Yamaguchi1997}. Very recent work has shown that a custom implementation of phase-shifting holography can be used to reconstruct the complex field of a twisted light mode \cite{DErrico2017}. While this method is functional, it has disadvantages: a time consuming circular-path integration of square-grid camera pixels, a separate reference path that can introduce additional phase noise, and intermediary optical transformations that raise the possibility of significantly increased error\cite{Dennis2012}.

In this paper, we show that standard phase-shifting digital holography  techniques \cite{Yamaguchi1997} can be applied to complex optical fields that carry OAM, characterizing them fully and quantifying their OAM power spectra with excellent fidelity. We use composite gratings on a spatial light modulator (SLM), which generates both the studied beam and a co-propagating, phase-controlled reference. We show that this complex field measurement is sufficient to decompose the field into a linear combination of Laguerre Gaussian (LG) modes. This is a fast, direct, and accurate determination of the OAM power spectrum, and the simplicity of our method minimizes error and uncertainty from alignment. Using this technique, we demonstrate an OAM measurement of the highest reported sensitivity, for both single and composite LG modes, and demonstrate that the measurement technique is robust against various sources of error.% of the highest purity vortex beams so far recorded.

% \textbf{Mark says: I would like to delete all of this. , and unlike the previous example of phase-shifting holography, we do not use intermediary optical transformation filters, vastly simplifying the implementation of the measurement and also reducing the possibility of optical misalignment effects. We also show that definition of the mode center of the beam is the primary source of error in this class of modal decomposition measurement, and this error can be mitigated by using a large beam and measurements of the mode center with sub-pixel accuracy.}

\section{Common-path phase-shifting digital holography of vortex beams}

In many interferometry experiments, the mode under test and the reference Gaussian overlap at an angle, leading to an additional linear phase that must be removed later. Independent beam paths also lead to the possibility of artificial fluctuations in the relative phase between the studied mode and the reference beam. This is the source of \textit{phase jitter} error. We address these problems by propagating the studied mode and the reference beam along a common path. To achieve this, we use a single hologram to generate both the beam being studied \textit{and} the Gaussian reference beam. This gives very precise control over the relative phase between the mode of interest and the reference mode, which minimizes error and uncertainty in the final, reconstructed phase.

\subsection{Phase-shifting Digital Holography and Hologram Generation}
%Paragraph GOAL: Describe PSDH process and why we want to use it

We begin by describing our implementation of phase-shifting digital holography: an extension of interferometry that enables phase and amplitude measurements of an arbitrary optical field \cite{Yamaguchi1997}. This technique requires the measurement of four interferograms with intensities $I(x,y,\phi_R)$ and relative phase $\phi_R$ between the studied mode and a Gaussian reference beam. The transverse phase of the desired mode, $\Phi (x,y)$, can then be calculated as 

\begin{equation}
\Phi (x,y)  =  tan^{-1}\left(\frac{I(x,y,\frac{3\pi}{2})-I(x,y,\frac{\pi}{2})}{I(x,y,0)-I(x,y,\pi)}\right).
\label{phase}
\end{equation}

\begin{figure}[!ht]
\centering
 \includegraphics[width=5in]{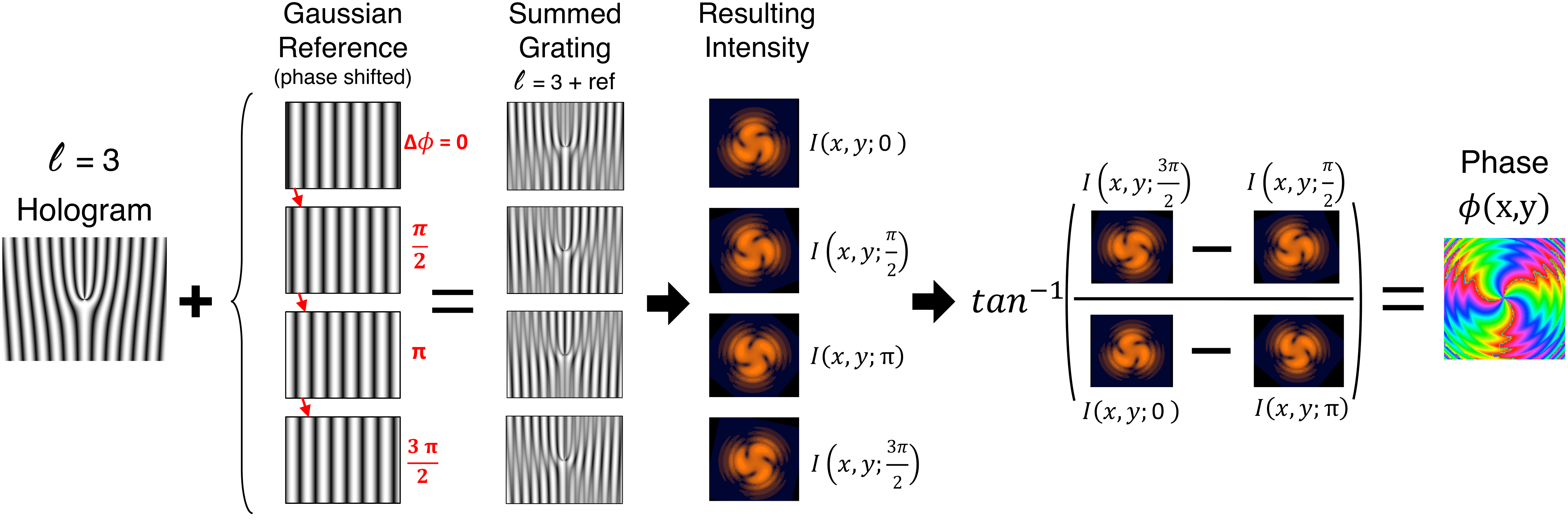}
 \caption{Schematic representation of phase-shifting digital holography }
  \label{prprocess}
\end{figure}

\noindent \cite{Yamaguchi1997} illustrated schematically in Fig. \ref{prprocess}.

%Paragraph GOAL: describe hologram generation and prprocess
To recover $\Phi(x,y)$, we begin by generating a series of four holograms. Each hologram is encoded with the sum of 1) the field of the studied mode superimposed with a plane wave, and 2) a phase reference grating such that amplitude of the hologram is given by 

\begin{equation}
A_H(x,y)= \frac{E_{studied}(x,y)}{E_{incident}(x,y)}* e^{\frac{2\pi x}{L}}+e^{\frac{2\pi x}{L}+\phi_R}
\label{ahologram}
\end{equation}

\noindent \cite{Kotlyar2007} where L is the grating constant. We normalize the studied mode by the experimentally measured field incident onto the SLM \cite{Clark2016}, and subsequently for each value of $\phi_R$, these holograms generate the interferograms needed for Eq. 1 in the first diffracted order. 

For the case of generating OAM, the plane wave is superimposed with a spiral phase, $e^{i \ell \phi}$, where $\ell$ is the OAM of the beam \cite{Clark2016}. Our measured incident field is a Gaussian that is then normalized out in the hologram. Without the last term in Eq. \ref{ahologram}, this creates a hologram such as that shown in the first column of Fig. \ref{phase}, which does not include a reference beam. The next column of Fig. 1 shows images of the last term in Eq. \ref{ahologram} that generate the reference beams at four phase steps from $0$ and $\frac{3\pi}{2}$ as is demanded by Eq. 1. Summing columns one and two (i.e. including both terms in Eq. 2) results in superposition holograms (column three) that each generate the desired mode with a phase-controlled reference beam along a common direction.

Light incident onto one of these holograms results in an interferogram, $I(x,y,\phi_R)$, in the first diffracted order. Each of the four interferograms can be recorded successively on a CCD before combining the images via Eq. \ref{phase} to calculate the phase of the desired mode at each pixel location.

%Paragraph GOAL: add how to recover amplitude
To obtain the complete optical field, we must also measure the field amplitude of the beam under test, which is produced by the hologram that contains no additional reference as in the first column of Fig.  \ref{prprocess}. This is recorded with a CCD, and by taking the square root of each pixel value, we obtain the amplitude of the mode. This allows for complete measurement of the complex amplitude and phase of the mode being studied.% in addition to comparison to other modes. 

%Paragraph GOAL: introduce different modes

Illuminating a standard forked diffraction grating, such as the first hologram in Fig. \ref{prprocess}, with a Gaussian beam results in a beam with pure OAM, but a superposition of Laguerre Gaussian (LG) radial modes. These beams are referred to as Hypergeometric Gaussian (HyGG) modes\cite{Kotlyar2007}. Pure LG modes can be produced via an additional amplitude mask within the hologram to match the amplitude of the desired LG mode\cite{Kotlyar2007}. Each of these LG and HyGG modes have unique phase profiles, even when they carry the same OAM. With this process, we can produce and measure these types of complex optical fields.

\subsection{Example of Phase-shifting Digital Holography with Experimental Data }
\begin{figure}[!ht]
\centering
 \includegraphics[width=5in]{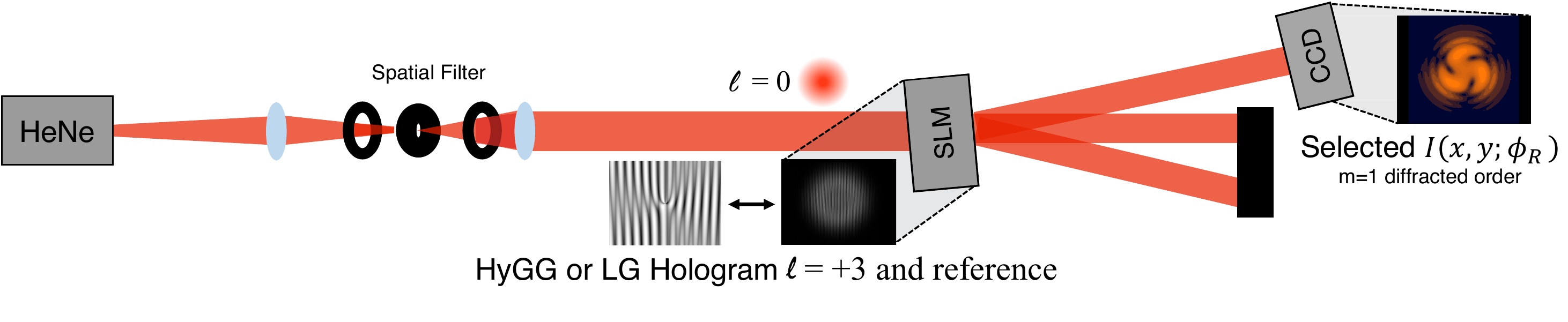}
 \caption{Experimental schematic: a HeNe laser passes through a spatial filter onto a SLM from which light in the first diffracted order is collected on a CCD. We project holograms with either HyGG or LG modes plus a phase controlled reference.}
  \label{prsetup}
\end{figure}

\begin{figure}[!ht]
\centering
 \includegraphics[width=4.5in]{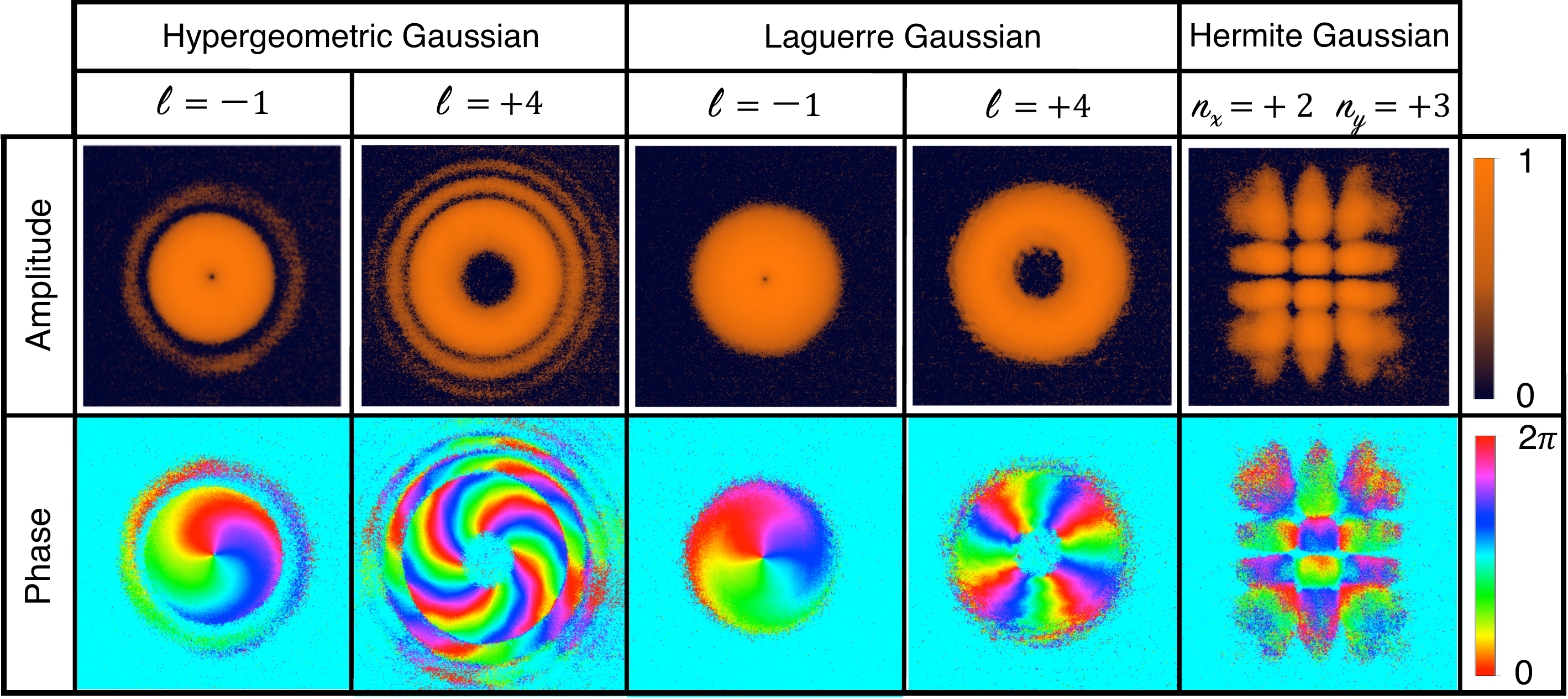}
 \caption{Experimental measurements of several complex laser modes; phase was measured by collinear phase-shifting digital holography.}
  \label{prresults}
\end{figure}

%Paragraph GOAL: Describe experimental set up
For experimental demonstration, a collimated and spatially filtered HeNe laser is passed through a transmission SLM, as shown in Fig. \ref{prsetup}. The light in the first diffracted order is measured by a Nikon D5200 camera at a distance of $0.15 z_R$ from the SLM. %(just for note in case we want to use it, zr is 22m).
All other diffracted orders are blocked. Fig. \ref{prresults} shows the results of the full complex measurements for HyGG modes with $\ell=+4$ and $\ell=-1$, of LG modes with $\ell=+4$ and $\ell=-1$, and of a Hermite Gaussian (HG) with $\mathit{n_x}=+2$ and $\mathit{n_y}=+3$. 

It is clear from our data that both the amplitude and the phase match expectations. In the amplitude measurements, we see smaller vortex cores for smaller $\ell$ values. We observe many radial modes in the HyGG beams, in contrast to the absence of radial modes in the LG beams. The phase information also shows a significantly higher amount of curvature in the HyGG modes as compared to the LG modes, as expected due to the additional radial modes present in HyGG beams \cite{Kotlyar2007}. The expected topological charge is clearly and directly seen in both the HyGG and LG beams. While a discussion of stability of this measurement is reserved for later, these complete measurements of complex fields gives us the ability to quantify the OAM spectrum in each beam.

%Paragraph GOAL: Transition into modal decomposition

\section{Modal Decomposition}
If a direct measurement of the amplitude and phase of the optical field is made, the field can be computationally decomposed into any basis, so that OAM and radial mode power spectra can be determined. This method has distinct advantages over the all-optical modal decomposition methods that have already been demonstrated \cite{DErrico2017}, including minimally required data acquisition (one complex beam profile is all that is needed to perform a computational modal decomposition, while an optical decomposition requires separate measurements for each mode), the ability to check other modal bases without acquiring more data, and minimal alignment error. Here we present the theory for modal decomposition in an LG basis, and demonstrate LG modal decomposition of our experimentally measured data shown in the previous section.

\subsection{Theory of Modal Decomposition}
We begin with an arbitrary complex scalar field $\Psi$, corresponding to that which can be measured through the techniques described above. This field can be expanded into the sum of Laguerre Gaussian components, so that 

\begin{equation}
\Psi=\sum_{l=-\infty}^{\infty} \sum_{p=0}^{\infty}C_p^{\ell}  LG_{p}^{\ell}.
\end{equation}
Orthogonality of LG modes allow for full characterization of the modal composition of $\Psi$ by way of

\begin{equation}
C_p^{\ell}=\int_{All}\Psi LG_{p}^{\ell*} dA.
\end{equation}

Given experimental measurements of the complex image of arbitrary field $\Psi$, we can measure $C_p^{\ell}$ by multiplying that image by a calculated image of $LG_p^{\ell *}$ and then by summing over all pixels. Thus after we have a complex image of a field, our method becomes a matter of iterating through digital transmission filters ($LG_p^{\ell *}$ for all relevant values of $\ell$,p) and summing over all pixels, which allows us to measure $C_p^{\ell}$ over a very large portion of the $\lbrace\ell, p\rbrace$ parameter space very quickly, without having to take additional, physical measurements for each $\ell$. 

Based the decomposition techniques demonstrated in this work, one might be tempted to attempt measuring the $\ell$ power spectrum more directly by using spiral phases alone as a basis for decomposition into an angular momentum distribution. We take a moment to make very clear that for an input field $\Psi$ with arbitrary radial distribution, one cannot measure the angular momentum distribution with spiral phase-only transmission filters. That is to say that in general 
\begin{equation}
\abs{\int_{All}\Psi e^{-i \ell \phi} dA}^2 \neq \sum_{p=0}^{\infty}\abs{\int_{All}\Psi  LG_{p}^{\ell*}  dA}^2 = \abs{C_{\ell}}^2.
\end{equation}
This can be understood conceptually as a statement of the non-uniformity of the radial decomposition of a plane wave. One can imagine a beam of the form $\psi = \frac{1}{\sqrt[]{2}}(LG_{p_1}^{\ell_1}+LG_{p_2}^{\ell_2})$, in which $p_1 \neq p_2$ and $\ell_1 \neq \ell_2$. The angular momentum spectrum of $\psi$ is clearly such that the power is split evenly between the $\ell_1$ and $\ell_2$ OAM states. However 
\begin{equation}
\abs{\int_{All}\psi e^{-i \ell \phi} dA}^2 =\abs{\frac{1}{\sqrt[]{2}}\int_{All}\right( \abs{ LG_{p_1}^{\ell_1}}e^{i \ell_1 \phi}+\abs{ LG_{p_2}^{\ell_2}}e^{i \ell_2 \phi}\left) e^{-i \ell \phi} dA}^2 = \abs{\frac{1}{\sqrt[]{2}}\delta_{\ell \ell_n}\int_{All}\abs{ LG_{p_n}^{\ell_n}}}^2
\end{equation}
Thus as the integral of the absolute value of $LG_{p_1}^{\ell_1}$ does not equal that of $LG_{p_2}^{\ell_2}$, the correct angular momentum distribution is not recovered.

\subsection{Example of Modal Decomposition with Experimental Data}
\begin{figure}[!ht]
\centering
 \includegraphics[width=5in]{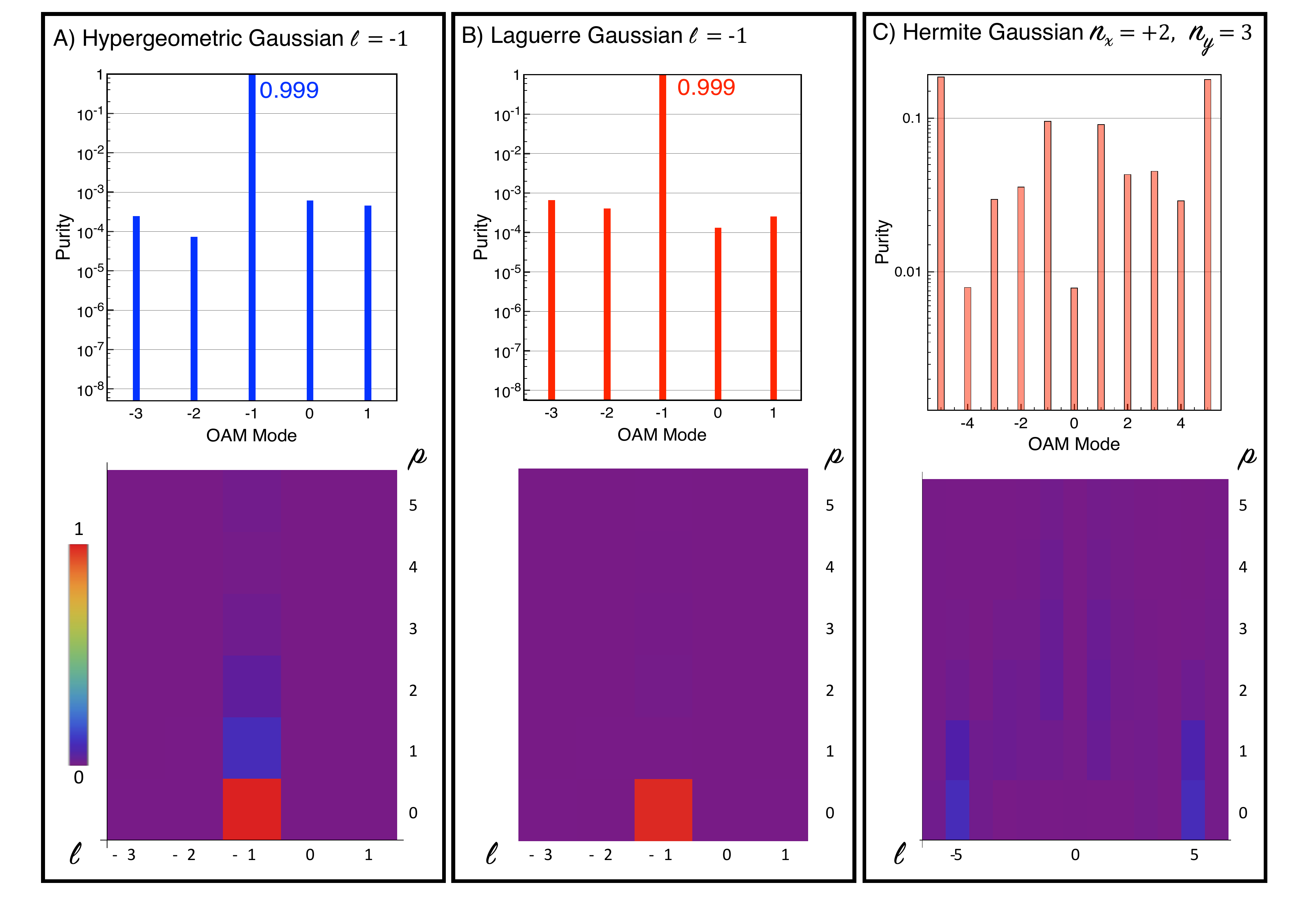}
 \caption{Experimental results for A) a HyGG beam of l=-1, B) an LG beam of l=4 and C) a HG beam of $n_x=2$ and $n_y=3$.  Note the symmetry in the spectrum for the HG mode demonstrating a net zero OAM in this mode. }
  \label{fulldecomp}
\end{figure} 
We demonstrate the use of our LG decomposition technique on three modes. The first is the HyGG with topological charge $\ell=-1$, which corresponds to the mode produced by the Gaussian illumination of a spiral phase optic with $2 \ell \pi$ phase wraps. The second and third are the well known $LG_{0}^{-1}$ and $HG_{n_x=2,n_y=3}$ modes. The results are shown below in Fig. \ref{fulldecomp}.

The modal decomposition reveals a full $\ell$-$p$ spectrum measurement. For the HyGG beam, we confirm in the 2D space a distribution of radial modes with most of the content in the $\ell =-1$ column. From this spectrum we project onto the the $\ell$ axis to get the OAM power spectrum. We find a purity in the fundamental OAM mode of $99.9 \%$. For our $LG^{-1}_0$ mode, we also find a purity of $99.9\%$.  The HG mode has a large radial mode distribution and is symmetric about $\ell=0$ giving this mode no net OAM, as expected.

\section{Error and Stability Analysis}
Successfully quantifying the errors affords us high confidence in our ability to accurately measure these OAM modes and confirm our high purities.

\subsection{Phase Stability Measurements}

Because our system is common path, we expect that the phase jitter errors are smaller than in an alternative system that measures interference from separate beam paths. To confirm this, we compare phase jitter in two different systems: our system used to produce OAM modes and a Michelson interferometer system. We observe stability in the phase profiles over time shown below in Fig. \ref{phasejitter}.

\begin{figure}[!ht]
\centering
 \includegraphics[width=4in]{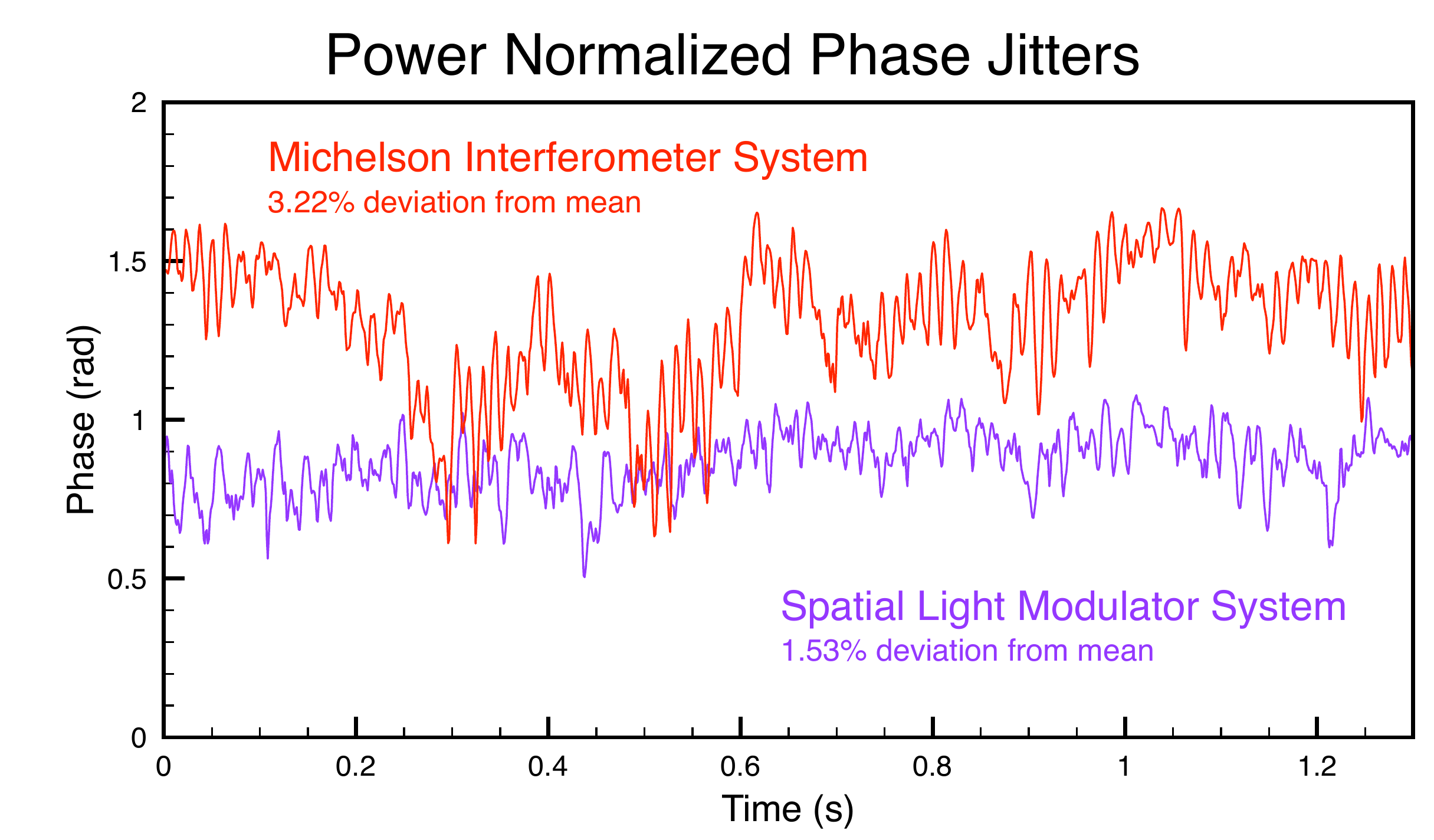}
 \caption{Phase jitter measurements of two different systems: our SLM system as compared to a Michelson Interferometer system. There is a noticeably larger fluctuation in the beam for the Michelson Interferometer as compared to the SLM system. When looking at the percent deviation from the mean value, we see a reduction in the phase jitter by a factor of 2.1. }
  \label{phasejitter}
\end{figure}

For the measurement, we record power fluctuations of small portions of interferograms using a photo diode and a pinhole. We can then convert this into a phase via $V_{diode}=V_{max} cos^2 \left (\frac{\phi_j}{2}\right)$, where $\phi_j$ is the phase between the beams. We note that the power fluctuations of any given beam will increase as it passes through an SLM. A polarizer can be placed before the SLM and rotated such that these fluctuations are minimized. We see that for the Michelson interferometer system, the deviation of the phase from the mean value is higher as compared to the SLM system, confirming that our technique has a larger degree of phase stability.

\subsection{Finite Window and Pixelation Error Analysis}
\begin{figure}[!ht]
\centering
 \includegraphics[width=4.5in]{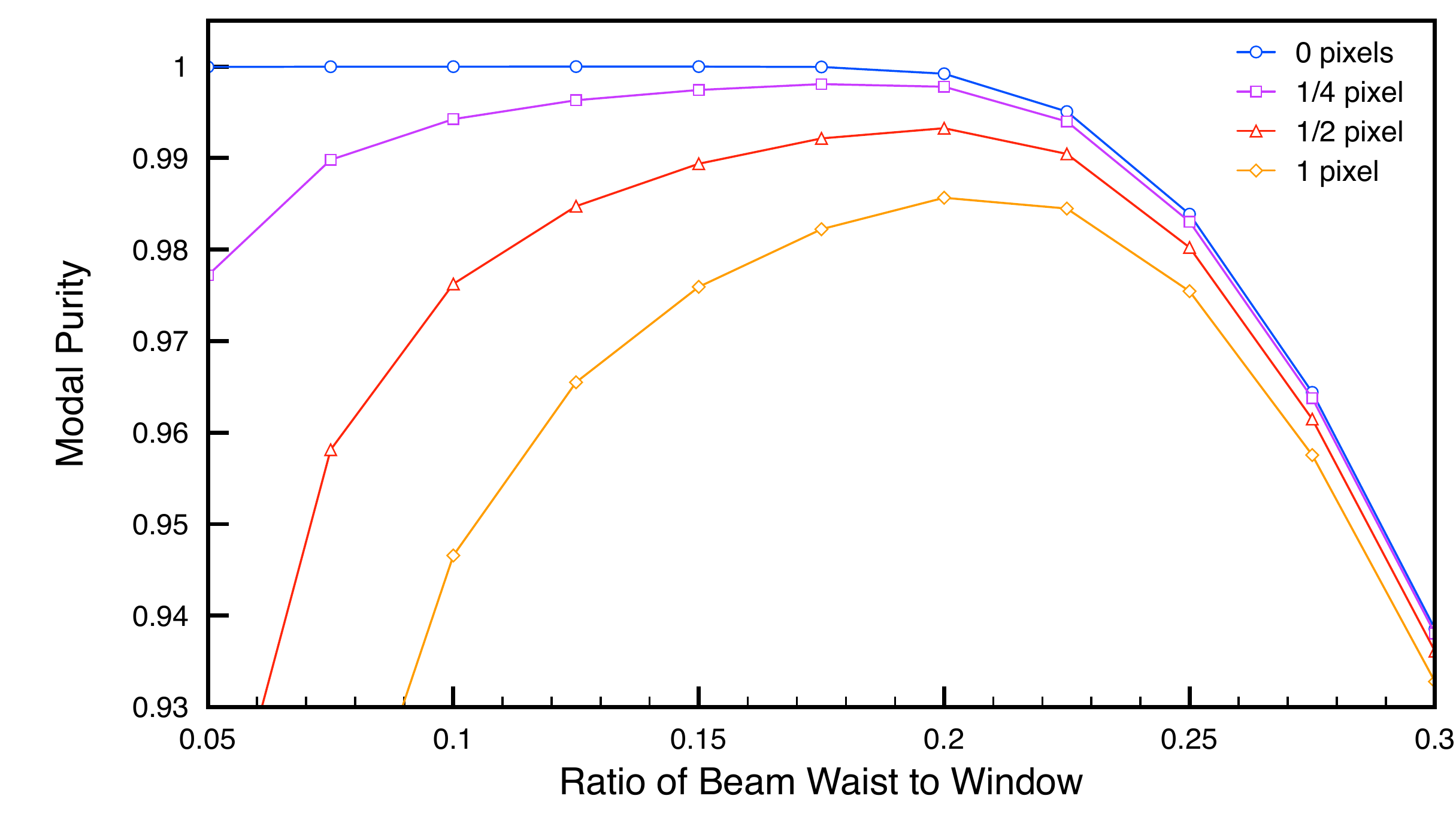}
 \caption{Modeled modal purity of a pure $LG^4_0$ mode as a function of beam waist to the total image window. Each curve shows the misalignment error resulting from the measured mode and calculated mode having: no relative displacement (blue), a quarter pixel displacement (purple), a half pixel displacement (red), one pixel displacement(orange). We conclude there is an optimal ratio for which the error is mostly negligible, and at which very small (sub-pixel) misalignment is at its most forgiving. This relative size is approximately $\frac{2}{5}$ of the window, for all values of $\ell$. We find slight variations in these errors for different OAM values, but the error-minimizing value of the beam to window ratio remains the same. }
  \label{err}
\end{figure}

As was described above in Section 3.1, the modal decomposition of a discretized complex field requires the generation of a set of transformation fields whose centers match that of the field of study. The alignment of the pixel edges of the generated transformation fields in respect to those of the measured image turns out to be extremely important. However, as the pixel edges are generally fixed for a physical measurement system, and as the alignment of pixel edges can take any place during computational analysis, we show in Fig. \ref{err} that if understood, errors from pixelation are very small even for low resolution images.

That the errors stemming from pixelation are not a function of resolution as much as they are a function of pixel-edge alignment is surprising, and is a result that can be exploited to achieve high spectral resolution from easy to compute, low resolution data.

Although even inexpensive consumer camera CCDs can be used to record complex images of resolutions on the order of thousands of pixels squared, here we choose to analyze modeled images with only 100 pixels squared to demonstrate that excellent results can be achieved with low resolution, rapidly computable data. We calculated the spectrum of pure, but discretized, LG modes using different relative displacements between the pixel edge of the modeled image and the transformation filter, and by changing the relative waist size of the mode with respect to the size of the image window. We found that the error in such a measurement is dependent on this relative beam size: too small and the pixel effects are less forgiving, but too large and the beam is clipped by the window. This trend is shown for $LG_{0}^4$ in Fig. \ref{err}, in which each line represents the measured modal purity of as a function of beam size in a fixed window, for different pixel-edge displacements.

\subsection{Detector Tilt Error Measurements}
\begin{figure}[!ht]
\centering
 \includegraphics[width=5in]{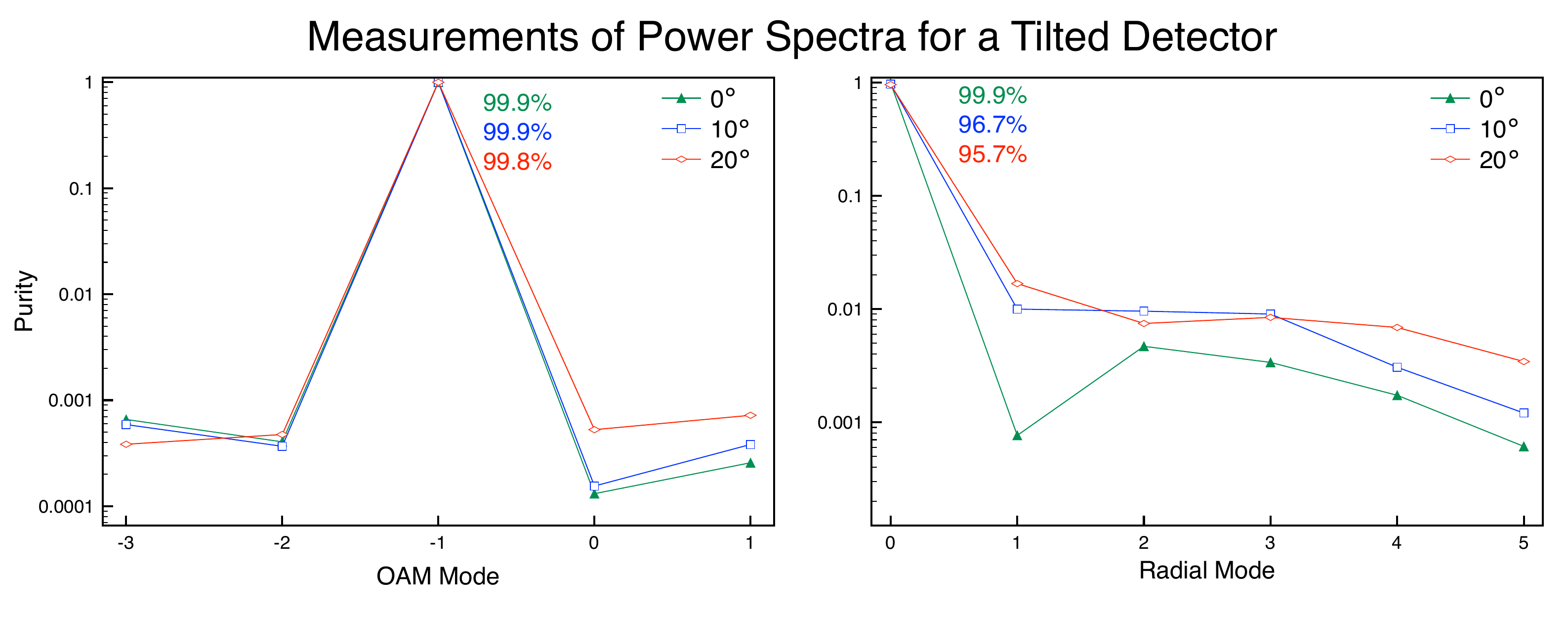}
 \caption{Measured spectra from CCD images taken at 0, 10 and 20 degrees of misalignment from the axis of propagation. }
  \label{tilt}
\end{figure}

Further measures could be taken into account such as correcting for any potential misalignments in the detector that is measuring the OAM\cite{Zhao2017}. However, in Fig. \ref{tilt}, we show that our technique is highly insensitive to the tilt of the detector. We see only a small deviation in the OAM power spectrum as we increase the tilt of the camera to 10 degrees. When increased to 20 degrees, we see slightly more power in the surrounding modes, but continue to measure a purity of 99.9\% in this case. The radial mode spectrum proves to follow the same pattern, but we do observe a decrease in the purity of the p=0 mode. That we observe high purity in the measured mode even in the case of dramatic detector misalignment is indicative of the strength of this technique.  

This makes some intuitive sense in that a tilted OAM beam may look oblong in the intensity measured on a camera, but because we also recover the phase, the OAM measurement is preserved. These complete error analyses combined with decomposition results demonstrate the robustness of the combined methods of phase-shifting digital holography and our digital modal decomposition.

\section{Conclusion}
In conclusion, we have demonstrated a novel collinear implementation of phase-shifting digital holography to measure the complex field of twisted light. This robust phase recovery allows for fast computational determination of the LG modal spectrum, and therefore the OAM power spectrum. We demonstrate how to mitigate sources of error in such a modal decomposition and find that using these methods we can reliably measure OAM modal purity up to 99.9\%.

\section*{Acknowledgements}
The authors acknowledge useful conversations with Brendan Heffernan, and financial support from the National Science Foundation (1509733, 1553905).  

%% \ackrule

%\section*{Biographies}

%\textbf{P. W. Wachulak} received the degree${\ldots}$ \\[6pt]
%\textbf{M. C. Marconi} received the degree${\ldots}$ \\[6pt]
%\textbf{R. A. Bartels} received the degree${\ldots}$ \\[6pt]
%\textbf{C. S. Menoni} received the degree${\ldots}$ \\[6pt]
%\textbf{J. J. Rocca} received the degree${\ldots}$

%\begin{equation}
%\begin{split}
%LG_{\ell,p}(r,\phi,z) =\sqrt{\frac{p!}{\left| l\right| !+p}} \frac{w_0}{w(z)}\left(\frac{\sqrt{2} \hspace{.5em} r}{w(z)}\right)^{\left| l\right| } L_p^{\left| l\right| }\left(\frac{2 r^2}{w(z)^2}\right)e^{i l \phi} \times \Phi (r,z)
%\end{split}
%\end{equation}
%\begin{equation}
%\Phi(r,z)= e^{-i (\left| l\right| +2 p+1)+\frac{i k_{0} z r^2}{2 \left(z^2+z_{R}^2\right)}-\frac{r^2}{w_{0}^2 \left(\frac{z^2}{z_{R}^2}+1\right)^2}}
%\label{PHI}
%\end{equation}
%(will keep fixing this ie gouy phase etc)
%\noindent where $w(z)= w_{0} \left(\frac{z^2}{z_{R}^2}+1\right)$

\end{document}